\documentclass[]{llncs}
\usepackage[ansinew]{inputenc}
\usepackage{epsfig}
\usepackage{amsfonts}
\usepackage{amssymb}
\usepackage{amsmath}
\usepackage{subfigure}
\usepackage{color}
\usepackage{comment}
\renewcommand{\tabcolsep}{10pt}

\newcommand{\textunderscript}[1]{$_{\text{#1}}$}

\newcommand{\RAW}{\text{RAW}}
\newcommand{\SBF}{\text{SBF}}
\newcommand{\DTIM}{\text{DTIM}}
\newcommand{\TIM}{\text{TIM}}
\newcommand{\DTIMIE}{\text{DTIM IE}}
\newcommand{\TIMIE}{\text{TIM IE}}
\newcommand{\RAWIE}{\text{RAW IE}}
\newcommand{\SIFS}{\text{SIFS}}
\newcommand{\DIFS}{\text{DIFS}}
\newcommand{\PSPOLL}{\text{PS\_ POLL}}
\newcommand{\RTS}{\text{RTS}}
\newcommand{\CTS}{\text{CTS}}
\newcommand{\ACK}{\text{ACK}}
\newcommand{\DATA}{\text{DATA}}
\newcommand{\UL}{\text{UL}}
\newcommand{\DL}{\text{DL}}
\newcommand{\MC}{\text{MC}}
\newcommand{\TIMO}{\text{TIMO}}
\newcommand{\noTIMO}{\text{non-TIMO}}
\newcommand{\STA}{\text{STA}}
\newcommand{\IE}{\text{IE}}

\begin{document}

\title{Capacity Analysis of IEEE 802.11ah WLANs for M2M Communications}
\author{T. Adame, A. Bel, B. Bellalta, J. Barcelo, J. Gonzalez, M. Oliver}
  
\institute{NeTS Research Group \\Universitat Pompeu Fabra, Barcelona\\
	    \emph{\{toni.adame, albert.bel, boris.bellalta, jaume.barcelo, javier.gonzalez, miquel.oliver\}@upf.edu}}

\date{October 2013}

\maketitle

\begin{abstract}
Focusing on the increasing market of the sensors and actuators networks, the IEEE 802.11ah Task Group is currently working on the standardization of a new amendment. This new amendment will operate at the sub-1GHz band, ensure transmission ranges up to 1 Km, data rates above 100 kbps and very low power operation. With IEEE 802.11ah, the WLANs will offer a solution for applications such as smart metering, plan automation, eHealth or surveillance. Moreover, thanks to a hierarchical signalling, the IEEE 802.11ah will be able to manage a higher number of stations (STAs) and improve the 802.11 Power Saving Mechanisms. In order to support a high number of STAs, two different signalling modes are proposed, TIM and Non-TIM Offset. In this paper we present a theoretical model to predict the maximum number of STAs supported by both modes depending on the traffic load and the data rate used. Moreover, the IEEE 802.11ah performance and energy consumption for both signalling modes and for different traffic patterns and data rates is evaluated. Results show that both modes achieve similar Packet Delivery Ratio values but the energy consumed with the TIM Offset is, in average, 11.7\% lower. 

\end{abstract}


\section{Introduction} \label{Sec:Intro}
In the last years, several draft amendments to IEEE 802.11 are being developed to support its growth into the future of wireless networking. These amendments seek to respond to the new needs of wireless communications, such as very-high throughput WLANs (IEEE 802.11ac \cite{Ong}, IEEE 802.11ad \cite{Cordeiro}), occupancy of TV Whitespaces (IEEE 802.11af \cite{Flores}) or sensor networks (IEEE 802.11ah \cite{aust2012ieee}), among others.

With respect to the Wireless Sensor Networks (WSNs), IEEE 802.11 is not suitable for applications based on this kind of devices as it was originally designed to offer high throughput to wireless communications without having into account energy consumption concerns. Many MAC protocols in general wireless ad-hoc networks assume more powerful radio hardware than the common one in sensor nodes, which is needed to run for months or years with just a pair of AA+ batteries \cite{Zhou}. Therefore, these particular constraints of sensors in terms of energy consumption require the design of new energy saving mechanisms which force them to remain asleep the maximum time possible during their operation periods.

In fact, the cost benefit offered by wireless sensors when compared with traditional wired sensors is inducing to predict that the Compound Annual Growth Rate (CAGR) of these systems could range 55\% to 130\% over the 2012-2016 time frame \cite{Harbor}. Excluding consumer short-range standards, this could amount to a market differential of as few as 300 million WSN connections in 2016 or as many as over 2 billion connections. 

The IEEE 802.11ah Task Group is nowadays working for enlarging the Wi-Fi applicability area, by designing a sub-1GHz protocol which will allow up to 8191 devices attached to a single Access Point (AP) to get access for short-data transmissions \cite{Draft802.11ah}. The standardization work was started in November 2010 and the final standard is expected not before January 2016, when it will be suitable for supporting Sensor Networks, Backhaul Networks for Sensors and Machine-to-Machine (M2M) Communication.

As for energy consumption, IEEE 802.11ah introduces new power saving features based on the segmentation of channel access into different contention periods, that are allocated to groups of stations according to a hierarchical distribution. Besides, ultra-low power consumption strategies are being developed from the former Power Saving Mode (PSM) of IEEE 802.11, so that they could extend up to 5 years the time that a STA can remain asleep without being disassociated from the network. Moreover, two signalling modes are defined in the draft amendment: the TIM Offset and Non-TIM Offset. While the first one uses two levels from the hierarchical distribution, has a high beacon transmission rate and sends little signalling information, the second one uses only one hierarchical level, has a low beacon rate and sends more signalling information. 

In this paper, the feasibility of IEEE 802.11ah WLANs to support a large number of stations with a low energy consumption is analyzed. A theoretical model of the network capacity, in terms of the maximum number of supported STAs, for the TIM Offset and Non-TIM Offset signalling modes defined in the draft amendment is presented. The presented results show that, for low traffic loads, the maximum number of stations defined in the draft amendment (8191) could be supported in non-saturated conditions if the data rate is higher than 2.4 Mbps. Besides, the low energy consumption of STAs is demonstrated from the comparison of the time spent in sleeping and non-sleeping states. As for the signalling modes, although Non-TIM Offset supports a slightly higher amount of stations without being saturated, TIM Offset offers a better global performance by saving up to 15\% of energy in high traffic scenarios.

The remainder of this paper is organized as follows: In Section \ref{Sec:IEEE80211ah}, the main features of the amendment in terms of PHY and MAC layer are described. A description of IEEE 802.11ah appears in Section \ref{Sec:MAC}, while the theoretical model about its capacity is proposed in Section \ref{capacity}. The results obtained in simulations, related to capacity and energy consumption, are shown in Section \ref{Sec:PEVA}. Finally, in Section \ref{Sec:Conclusions}, we present our conclusions and propose future work.


\section{IEEE 802.11ah} \label{Sec:IEEE80211ah}

IEEE 802.11ah is being designed for supporting applications with the following requirements \cite{aust2012sub}: up to 8191 devices associated to an AP, adoption of Power Saving strategies, minimum network data rate of 100 kbps, operating carrier frequencies around 900 MHz, coverage up to 1 km in outdoor areas, one-hop network topology and short and infrequent data transmissions (data packets $\sim$ 100 bytes).

One of the goals of the IEEE 802.11ah Task Group (TGah) is to offer a standard that, apart from satisfying these previously mentioned requirements, minimizes the changes with respect to the widely adopted IEEE 802.11. In that sense, the proposed PHY and MAC layers are based on the IEEE 802.11ac standard and moreover, try to achieve an efficiency gain by reducing some control/management frames and the MAC header length. 

Technologies like Orthogonal Frequency Division Multiplexing (OFDM), Multi Input Multi Output (MIMO) and Downlink Multi-User MIMO (DL MU-MIMO) - which was firstly introduced in the IEEE 802.11ac - are also employed by the 802.11ah system.

\subsection{PHY Layer}

\subsubsection{Channelization}

As commented above, the IEEE 802.11ah operates at the sub-1 GHz band, by being a 10 times down-clocked version of IEEE 802.11ac. This new standard defines different channel widths: 1 MHz, 2 MHz, 4 MHz, 8 MHz and 16 MHz.  

The available sub 1 GHz ISM bands differ depending on the country regulations, so that the IEEE 802.11ah has defined the channelization based on the wireless spectrum in different countries \cite{Hazmi2012}.

\subsubsection{Transmission Modes}

The common channels adopted by the IEEE 802.11ah are 2 MHz and 1 MHz. Hence, the PHY layer design can be classified into 2 categories: transmission modes greater or equal than 2 MHz channel bandwidth and a transmission mode of 1 MHz channel bandwidth.  

For the first case, the PHY layer is designed based on 10 times down-clocking of IEEE 802.11ac's PHY layer; i.e. the PHY layer uses an OFDM waveform with a total of 64 tones/sub-carriers (including tones allocated as pilot, guard and DC), which are spaced by 31.25 kHz. The modulations supported include BPSK, QPSK and 16 to 256 QAM (Table \ref{tab:data_rate}). It will also support multi user MIMO and single user beam forming. For the second case, the tone spacing is maintained, but the waveform is formed with 32 tones, instead of 64 \cite{m2mwordpress}. 

\begin{table}
\begin{center}
\begin{tabular}{|c|c|c|c|}
\hline 
\textbf{MCS Idx} & \textbf{Mod} & \textbf{R} & \textbf{Data Rate (kbps)} \\ 
\hline 
0 & BPSK & 1/2 & 300 \\ 
\hline 
1 & QPSK & 1/2 & 600 \\ 
\hline 
2 & QPSK & 3/4 & 900 \\ 
\hline 
3 & 16-QAM & 1/2 & 1200 \\ 
\hline 
4 & 16-QAM & 3/4 & 1800 \\ 
\hline 
5 & 64-QAM & 2/3 & 2400 \\ 
\hline 
6 & 64-QAM & 3/4 & 2700 \\ 
\hline 
7 & 64-QAM & 5/6 & 3000 \\ 
\hline 
8 & 256-QAM & 3/4 & 3600 \\ 
\hline 
9 & 256-QAM & 5/6 & 4000 \\ 
\hline 
10 & BPSK & 1/4 & 150 \\ 
\hline 
\end{tabular}
\vspace{5mm}
\caption{IEEE 802.11ah Modulation and Codification Schemes and their corresponding Coding and  Data Rates for Bandwidth (BW) = 1MHz, and Number of Spatial Streams (NSS) = 1 \cite{Draft802.11ah}}
\label{tab:data_rate}
\end{center} 
\end{table}

\subsection{MAC Layer}

\subsubsection{Hierarchical Grouping}

IEEE 802.11 MAC layer defines that the AP assigns an Association IDentifier (AID) to each STA. The maximum number of stations mapped is only 2007, due to the length of the partial virtual bitmap of Traffic Indication Map (TIM) Information Element (IE).

In order to support a larger number of STAs, TGah has defined a novel and hierarchical distribution of them \cite{TGah}. With this novel structure, the IEEE 802.11ah achieves the objective of supporting up to 8191 STAs. In this manner, the AIDs are classified into pages, blocks (from now on called \textit{TIM Groups} in this paper), sub-blocks and STAs' indexes in sub-blocks.

\begin{figure}[h!]
\centering
\includegraphics[scale=0.35]{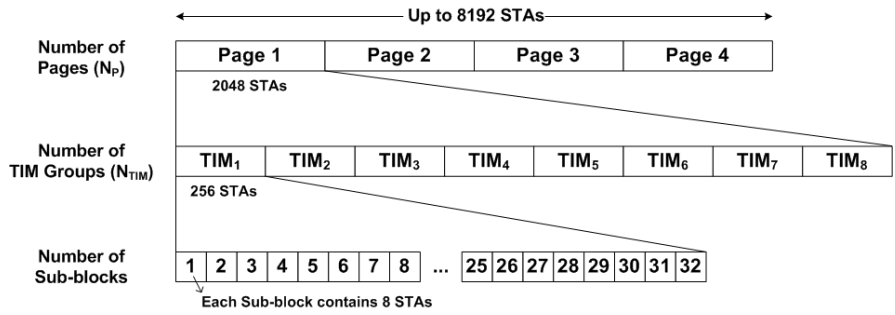}
\caption{Hierarchical distribution of stations in an IEEE 802.11 network}
\label{fig:hierarchical}
\end{figure}

The number of pages $(N_{P})$ and TIM Groups per page $(N_{\TIM})$ is configurable according to the size and requirements of the network. An example of hierarchical distribution with $N_{P}=4$ and $N_{\TIM}=8$ is shown in Figure \ref{fig:hierarchical}.

\begin{figure}[h!]
\centering
\includegraphics[scale=0.35]{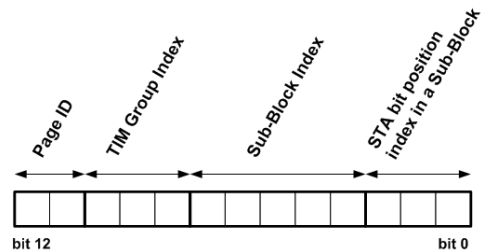}
\caption{AID Structure}
\label{fig:AID}
\end{figure}

\begin{figure}[h!]
\centering
\includegraphics[scale=0.35]{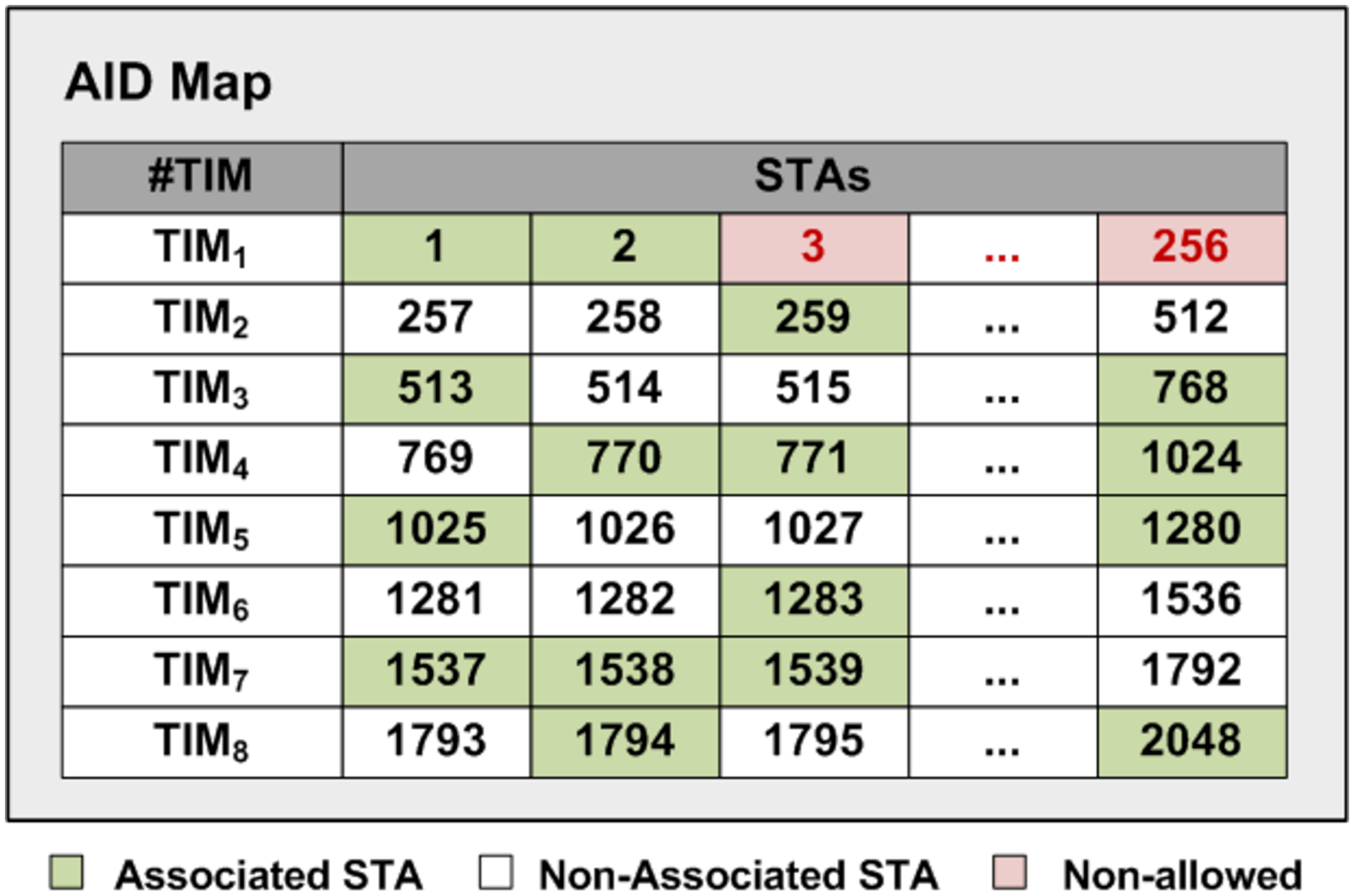}
\caption{Example of AID Map}
\label{fig:AIDmap}
\end{figure}

The IEEE 802.11ah AID assigned is unique and consists of 13 bits (see Figure \ref{fig:AID}) that include the different hierarchical levels. It could be an effective way to categorize STAs with respect to their type of application, battery level or required QoS (Figure \ref{fig:AIDmap}). Thus, QoS differentiation could be achieved by restricting the number of stations in high-priority groups, in order to limit the contention in them.

\subsubsection{Beacon Structure}

There are two classes of signalling beacons. The first one, which is called Delivery Traffic Indication Map (DTIM), informs about which groups of STAs have pending data at the AP and also about multicast and broadcast messages. The second class of beacons is called simply TIM. Each TIM informs a group of STAs about which of them have pending data in the AP.

Both DTIM and TIM beacon structures (Figure \ref{fig:beacons}) are based in one Short Beacon Frame plus optional Information Elements (IE) for different purposes:

\begin{figure}[h!]
\centering
\includegraphics[scale=0.5]{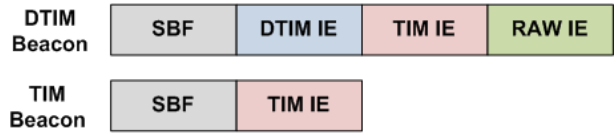}
\caption{DTIM and TIM Beacon structure}
\label{fig:beacons}
\end{figure}

\begin{itemize}
\item SBF (Short Beacon Frame):
Its primary functions are: advertizing the AP presence and synchronizing the STAs.

\item DTIM IE:
It is only transmitted in DTIM beacon frames and not in TIM segments. From this element, STAs can deduce their assignment in TIM Groups and their wake-up intervals. Besides, STAs with their TIM Group bit set to 0 may not wake up at assigned TIM Group interval.

\item TIM IE:
When the complete traffic indication bitmap is divided into multiple TIM Groups, each TIM IE indicates which stations from its corresponding TIM Group have pending data to receive.

\item RAW IE (Restricted Access Window IE):
It is responsible for signalling all information related to RAW; i.e. the time period in which selected STAs contend for accessing the channel. This IE includes: time from the beacon to the RAW, duration of the RAW as well as mechanisms to generate sub-slots within the RAW contention period.

\end{itemize}

\subsubsection{Types of Stations}

As defined in IEEE 802.11ah draft, there are 3 different kinds of STAs, each with its procedures and time periods to access the channel:

\begin{enumerate}
\item \textit{TIM Stations:}
They have to listen to both DTIM and TIM beacons to send or receive data.

\item \textit{Non-TIM Stations:}
They only have to listen to DTIM beacons to send or receive data.

\item \textit{Unscheduled Stations:}
These STAs do not need to listen to beacons and can transmit data anytime.

\end{enumerate}


\section{Power Saving Mechanisms in IEEE 802.11ah} \label{Sec:MAC}

\subsection{TIM and Page Segmentation} 

Compared to IEEE 802.11 power management mode, IEEE 802.11ah amendment tries to improve that power saving mode by means of using a scheme called \textit{TIM and page segmentation}. This new scheme aims to save STA's energy consumption not only when they do not have to send or receive any data, but also during their operation time, by allocating them in shorter contention periods with other few STAs. To achieve this goal, IEEE 802.11ah extends some of the mechanisms already introduced in the former IEEE 802.11 PSM version \cite{IEEE802.11-2012}.


IEEE 802.11 PSM is based on the inclusion of an IE field in each TIM beacon, responsible for signalling the existence of packets in the downlink buffer for each STA associated to the  AP. Thus, any node can enter into a power saving state if it observes in the TIM beacon that there is no downlink traffic aimed at it.

However, this mechanism has two major drawbacks:
\begin{itemize}
\item Firstly, all STAs in power saving mode are forced to listen to all TIMs and, therefore, to shorten their sleeping periods.
\item Secondly, each TIM must be able to map all STAs in the network (they could be up to 2048), so in a densely populated network such mapping
would be very long in size and expensive in terms of energy.
\end{itemize}

TGah, trying to overcome the drawbacks mentioned above, proposes a scheme based on hierarchical signalling. This hierarchy is reflected both in the organization of the STAs in groups and in the signalling beacons. Therefore, the STAs only remain active during the time assigned to their group, the signalling data is shorter than the former PSM and the network can manage a higher number of STAs.

Between two consecutive DTIMs, an AP broadcasts as many TIMs per page as groups of STAs. Each one of these TIMs informs STAs about buffered downlink packets. These packets will be dropped after a certain time determined by the size of the AP buffer and the association parameters chosen by the STA. 

\subsubsection{TIM Offset}
\label{TIMoffset}
The draft specification also includes the \textit{TIM Offset}, a 5-bit field that is contained in the DTIM IE and allows the AP to indicate the TIM beacon offset with respect to the DTIM beacon. The corresponding TIM beacon for the first TIM Group of a specific page can be allocated at the indicated TIM offset. Thus, TIM Groups of different pages can be flexibly scheduled over beacon intervals.

\begin{figure}[h!]
\centering
\includegraphics[scale=0.35]{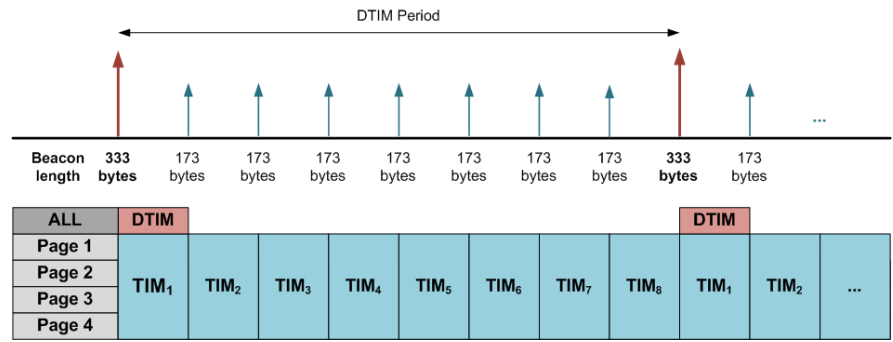}
\caption{Beacon distribution using Non-TIM Offset signalling mode in a four-page IEEE 802.11ah network}
\label{fig:offset1}
\end{figure}

\begin{figure}[h!]
\centering
\includegraphics[scale=0.35]{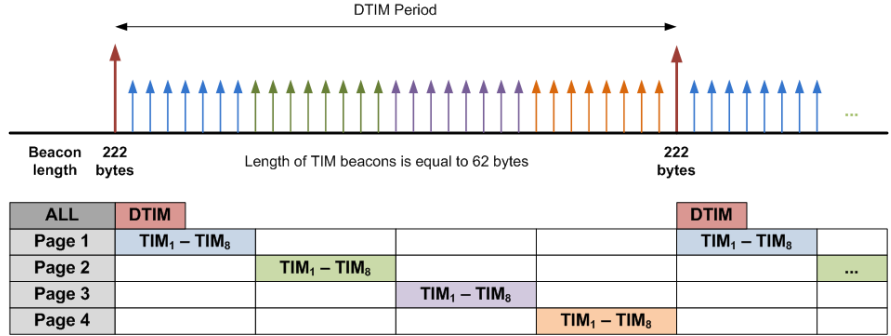}
\caption{Beacon distribution using TIM Offset signalling mode in a four-page IEEE 802.11ah network}
\label{fig:offset2}
\end{figure}

If TIM Offset is not used (Figure \ref{fig:offset1}), signalling information of a determined TIM Group is transmitted in the same beacon as many times as existing pages in the network. This fact implies that the STAs are forced to listen the information related to pages that they do not belong to.

Otherwise, with the use of TIM Offset (Figure \ref{fig:offset2}), all beacons except DTIM contain signalling information addressed to STAs from a single page, reducing in this manner the time spent in the receiving state and, consequently, the energy consumed by STAs. In this mode, TIM beacons are sequentially sent from the first to the last page with a rate $N_{P}$ times higher than in the Non-TIM Offset case.

Length of DTIM and TIM beacons becomes an important parameter in the network, as the more pages it has, the more bits it will need to map all contained stations. The equations of the length of both beacons are shown in Table \ref{tab:par_beacon}, with different $L_{\TIM\;IE}$ values depending on the activation of the TIM Offset field.
\vspace{-0.7cm}

\begin{center}
\begin{align*}
L_{\DTIM} & = L_{\SBF}+L_{\DTIMIE}+L_{\TIMIE}+L_{\RAWIE}\\
L_{\TIM} & = L_{\SBF}+L_{\TIMIE}
\end{align*}
\end{center}
\vspace{-0.2cm}

\begin{table}
\begin{center}
\begin{tabular}{|c|c|c|}
\hline 
\textbf{L (bits)} & \textbf{TIM Offset} & \textbf{Non-TIM Offset} \\ 
\hline 
$L_{\SBF}$ & $200$ & $200$\\ 
\hline 
$L_{\DTIM\;\IE}$ & $(32+N_{\TIM}+N_{\TIM})\cdot N_{P}$ & $(32+N_{\TIM}+N_{\TIM})\cdot N_{P}$\\ 
\hline 
$L_{\TIM\;\IE}$ & $(40+2048/N_{\TIM})$ & $(40+2048/N_{\TIM})\cdot N_{P}$\\ 
\hline 
$L_{\RAW\;\IE}$ & $(16+N_{\TIM}\cdot 32)\cdot N_{P}$ & $(16+N_{\TIM}\cdot 32)\cdot N_{P}$ \\ 
\hline 
\end{tabular}
\vspace{5mm}
\caption{Length of beacon parameters according to the different signalling modes} 
\label{tab:par_beacon}
\end{center}
\end{table}
\vspace{-0.8cm}
\subsubsection{Channel Access}
\begin{figure}
\centering
\includegraphics[scale=0.4]{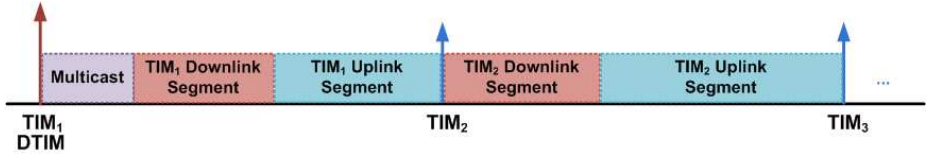}
\caption{Distribution of channel access into downlink and uplink segments}
\label{fig:segments}
\end{figure}

Once a node associates to an AP, it is included in a TIM Group and in its corresponding Multicast distribution group along with the other TIM Group stations. Figure \ref{fig:segments} shows how time between 2 consecutive TIMs is split into one Downlink (DL) segment, one Uplink (UL) segment as well as one Multicast (MC) segment placed immediately after each DTIM beacon. In our proposal, the proportion between $\psi \in \lbrace \DL,\UL\rbrace$ segments size is equal to the DL/UL traffic proportion $(\beta_{\psi})$, and we assume that the multicast segment is able to accommodate only one data packet.

The operation modes for the downlink and uplink cases are detailed below:
\begin{itemize}
\item Downlink:
When an AP needs to send a packet to a STA, the DTIM beacon has to include the TIM Group to which belongs that STA in its bitmap. Similarly, the corresponding TIM beacon has to include that STA, also, in its bitmap. Each signalled STA has to listen its TIM to know when to contend. This contention will be done using the Distributed Coordination Function (DCF), by sending first a PS-Poll frame, in order to get its corresponding data.

\item Uplink:
When a STA has to send an uplink message to the AP, it must listen its corresponding TIM Group for knowing when to contend the channel. In this case, the contention is also done through an DCF scheme. Both Basic Access (BA) and RTS/CTS mechanisms can be used.

\end{itemize}

\subsection{Long Sleeping Periods}

Apart from the \textit{TIM and Page Segmentation} scheme, an important feature in terms of energy savings of IEEE 802.11ah is the ability to set longer doze times (up to years) to STAs than IEEE 802.11. This is achieved by extending several system parameters during the initial handshake between an AP and its associated STAs.

However, an important drawback that has to be considered is the corresponding clock drift produced by such long doze times. Thus, the  more time a STA has been asleep, the further in advance it should wake up to avoid possible synchronization problems with the network \cite{IEEE802.11-12/1101r1}.

\section{Maximum number of supported STAs}
\label{capacity}

From the study of the channel access features in an IEEE 802.11ah WLAN, a theoretical model of the channel capacity for both the TIM Offset and Non-TIM Offset signalling modes is developed. The variables used to build the model are shown in Table \ref{tab:teo_parameters}.

\begin{table}[h!]
\begin{center}
\begin{tabular}{|c|l|c|}
\hline 
\textbf{Variable} & \textbf{Description} & \textbf{Unit} \\ 
\hline 
$T$ & DTIM Period & s \\ 
\hline 
$N_{\TIM}$ & Number of TIM Groups & - \\ 
\hline 
$N_{P}$ & Number of Pages & - \\ 
\hline 
$M_{\omega}$ & Signalling mode Scale factor & - \\ 
\hline 
$r$ & Network Data Rate & bps \\ 
\hline 
$CW_{\min}$ & Size of minimum contention window & slots \\ 
\hline 
$t_{\text{slot}}$ & Duration of an IEEE 802.11ah time slot & s \\ 
\hline 
$\alpha_{\psi}$ & Traffic Pattern for DL/UL & - \\ 
\hline 
$\beta_{\psi}$ & Proportion of DL/UL traffic & - \\ 
\hline 
$T_{\DTIM}$ & \vtop{\hbox{\strut DTIM Beacon time}\hbox{\strut $T_{DTIM} = \frac{L_{\DTIM}}{r}$}} & s \\ 
\hline 
$T_{\TIM}$ & \vtop{\hbox{\strut TIM Beacon time}\hbox{\strut $T_{DTIM} = \frac{L_{\TIM}}{r}$}} & s \\ 
\hline 
$T_{\psi}$ & Duration of a packet transmission & s \\ 
\hline 
$T_{\MC}$ & \vtop{\hbox{\strut Duration of a multicast packet transmission}\hbox{\strut $T_{MC}=\frac{L_{\DATA}}{r} + T_{\DIFS}$}} & s  \\ 
\hline 
$T_{\DL}$ & \vtop{\hbox{\strut Duration of a DL packet transmission}\hbox{\strut $T_{DL}=\frac{L_{\PSPOLL}}{r} + T_{\SIFS}+ \frac{L_{\DATA}}{r}+T_{\SIFS}+$}
\hbox{\strut $\;\;\;\;\;\;\;\;\;+\frac{L_{\ACK}}{r}+ T_{\DIFS}$}} & s \\ 
\hline 
$T_{\UL}$ & \vtop{\hbox{\strut Duration of an UL packet transmission}\hbox{\strut $T_{\UL}=\frac{L_{\RTS}}{r}+T_{\SIFS}+\frac{L_{\CTS}}{r}+T_{\SIFS}+$}
\hbox{\strut $\;\;\;\;\;\;\;\;\;+\frac{L_{\DATA}}{r}+T_{\SIFS}+\frac{L_{\ACK}}{r}+T_{\DIFS}$}} & s \\
\hline 
\end{tabular} 
\vspace{5mm}
\caption{List of parameters used in the IEEE 802.11ah capacity calculation}
\label{tab:teo_parameters}
\end{center}
\end{table}

The maximum number of packets in a DTIM period $(N_{\psi})$ is obtained from an equation formed by two summands: one corresponding to the DL/UL packets contained in the first TIM period $(N_{\psi,\DTIM})$ and the other corresponding to those contained in the rest of TIM periods $(N_{\psi,\TIM})$. 

The value of $N_{\psi}$ also depends on the $\omega \in \lbrace \TIMO,\noTIMO\rbrace$ signalling mode chosen (TIM Offset or Non-TIM Offset, respectively) through a scale factor $M_{\omega}$, that has been defined as $M_{\TIMO}=N_{\TIM} \cdot N_{P}$ for the TIM Offset case and as $M_{\noTIMO}=N_{\TIM}$ for the Non-TIM Offset case. 

\begin{center}
$ N_{\psi}=N_{\psi,\DTIM}+\left(M_{\omega}-1\right) \cdot N_{\psi,\TIM}$\\
\end{center}

In the theoretical model, $N_{\psi,\DTIM}$ and $N_{\psi,\TIM}$ are calculated dividing the corresponding DL/UL segment time by the duration of a DL/UL packet transmission, and finally taking the integer part. Due to the allocation of a multicast transmission in the first TIM period, $N_{\psi,\DTIM}$ is always lower than $N_{\psi,\TIM}$. In order to obtain a conservative result, the time corresponding to a whole backoff $CW_{min} \cdot t_{\text{slot}}$ has also been included in the DL/UL segments. 

\begin{center}
$N_{\psi,\DTIM}=\left\lfloor{\frac{\left(\frac{T}{M_{\omega}}-T_{\MC}-T_{\DTIM}\right) \cdot \beta_{\psi} - CW_{\min} \cdot t_{\text{slot}}} {T_{\psi}}} \right\rfloor$  \\

$N_{\psi,\TIM}=\left\lfloor{\frac{\left(\frac{T}{M_{\omega}}-T_{\TIM}\right) \cdot \beta_{\psi} - CW_{\min} \cdot t_{\text{slot}}}{T_{\psi}}} \right\rfloor $\\
\end{center}

To obtain the maximum number of stations in an IEEE 802.11ah WLAN, we assume that a STA is only capable of receiving and transmitting one data packet per DTIM interval (two packets in total). Besides, the network traffic pattern $(\alpha_{\psi})$ has to be taken into account. This value, between 0 and 1, represents the proportion of stations that have data to receive from or to transmit to the AP. Once applied this proportion to $N_{\psi}$, the resulting minimum between both operators corresponds to the maximum number of supported stations by the network.

\begin{center}
$ N_{\STA}=min \left(\frac{\displaystyle N_{\DL}}{\displaystyle \alpha_{\DL}},\frac{\displaystyle N_{\UL}}{\displaystyle \alpha_{\UL}} \right)$
\end{center}


\section{Performance Evaluation} \label{Sec:PEVA}

We simulate a fully connected IEEE 802.11ah WLAN in MATLAB with different number of scheduled TIM STAs, where packets are delivered from the source to the destination in just one hop and there are no hidden terminals. We also assume ideal channel conditions, without communication errors, delays or capture effects. It is considered that the AP and all STAs have infinite buffers, although a packet could be dropped if it is retransmitted, inside the same segment, more than R\textunderscript{max} times. 

The IEEE 802.11ah defines four different power states for the STAs: receiving, idle, transmitting and sleeping. We consider that STAs are only capable of receiving and transmitting one data packet per DTIM interval. These intervals have been split into 8 TIMs $(N_{\TIM})$ and also into 4 pages ($N_P$). We have not considered the presence of Non-TIM or Unscheduled STAs. 

The parameters considered in the different simulations are presented in Table \ref{tab:sim_param}, where $N_{\DTIM}$ corresponds to the simulation duration, in number of DTIMs. The list of Modulation and Codification Schemes used, as well as their corresponding data rates, appears in Table \ref{tab:data_rate}. 

\begin{table}[h!]
\begin{center}
\begin{tabular}{|l|r|l|r|l|r|}
\hline 
T\textunderscript{DTIM} & $1.6\; s$ & T\textunderscript{SIFS} & $16 \mu s$ & L\textunderscript{DATA} & 100 bytes \\
\hline
N\textunderscript{TIM} & $8$ & T\textunderscript{DIFS} & $34 \mu s$ & L\textunderscript{PS-POLL} & 14 bytes \\
\hline
N\textunderscript{DTIM} & $100$ & CW\textunderscript{min} & $16$ & L\textunderscript{ACK} & 14 bytes \\
\hline
N\textunderscript{P} & 4 & CW\textunderscript{max} & $1024$ & L\textunderscript{RTS} & 20 bytes \\
\hline
T\textunderscript{slot} & $9\; \mu s$ & R\textunderscript{max} & $7$ & L\textunderscript{CTS} & 14 bytes \\
\hline
\end{tabular}
\vspace{5mm}
\caption{List of Simulation Parameters}
\label{tab:sim_param}
\end{center}
\end{table}

We have considered three different scenarios (Table \ref{tab:sim_traffic}). At every DTIM, depending on the traffic pattern, only a percentage of the STAs will have a message from/to the AP; i.e., the AP generates a data message addressed to a percentage of randomly selected STAs $(\alpha_{\DL})$. Similarly, a fraction of randomly selected STAs generate a data message addressed to the AP $(\alpha_{\UL})$. From that information, the proportion of downlink/uplink traffic in our network, $\beta_{\DL}=\frac{\alpha_{\DL}}{\alpha_{\DL}+\alpha_{\UL}}$ and $\beta_{\UL}=1-\beta_{\DL}$ respectively, and the data generation rate per station $(\lambda_{\psi})$ can be determined.

\begin{table}[h!]
\begin{center}
\begin{tabular}{|l|c|c|c|c|c|c|}
\hline 
 & \multicolumn{2}{|c|}{\% STAs} & \multicolumn{2}{|c|}{\% Traffic} & \multicolumn{2}{|c|}{Data gen. rate / STA} \\
\hline 
 & $\alpha_{\DL}$ & $\alpha_{\UL}$ & $\beta_{\DL}$ & $\beta_{\UL}$ & $\lambda_{\DL}$ & $\lambda_{\UL}$ \\ 
\hline 
Scenario A & 15\% & 15\% & 50\% & 50\% & 75 bps & 75 bps \\ 
\hline 
Scenario B & 15\% & 30\% & 33.3\% & 66.6\% & 75 bps & 150 bps \\ 
\hline 
Scenario C & 15\% & 45\% & 25\% & 75\% & 75 bps & 225 bps \\ 
\hline 
\end{tabular} 
\vspace{5mm}
\caption{Traffic Patterns Definition}
\label{tab:sim_traffic}
\end{center}
\end{table}

In order to evaluate the capacity of IEEE 802.11ah WLANs using both the TIM Offset and Non TIM Offset signalling modes (see Subsection \ref{TIMoffset}), we present several theoretical and simulated results for different traffic patterns by means of analyzing different figures of merit.

\begin{itemize}
\item Firstly, using the theoretical model presented in Section \ref{capacity}, we evaluate the maximum number of STAs that the network is capable of supporting (i.e., point at which the network is not able to deliver all generated packets) for the two signalling modes.
\item Secondly, in order to provide more insights on the IEEE 802.11ah WLAN operation with a large number of STAs, the performance of the network in terms of PDR (Packet Delivery Ratio) and $\eta$ (Network Efficiency) has been evaluated by simulation. In this manner, the influence of important effects like the different density of STAs within TIM Groups or packet collisions can be studied.
\item Finally, the feasibility of both signalling modes in terms of energy consumption has also been studied.
\end{itemize}

\subsection{Maximum Number of STAs}

First of all, using the model presented in Section \ref{capacity}, we find the maximum number of STAs supported by an IEEE 802.11ah WLAN. The parameters considered are shown in Table \ref{tab:sim_param}.

\begin{center}
\begin{figure}[h!]
\centering
\includegraphics[scale=0.34]{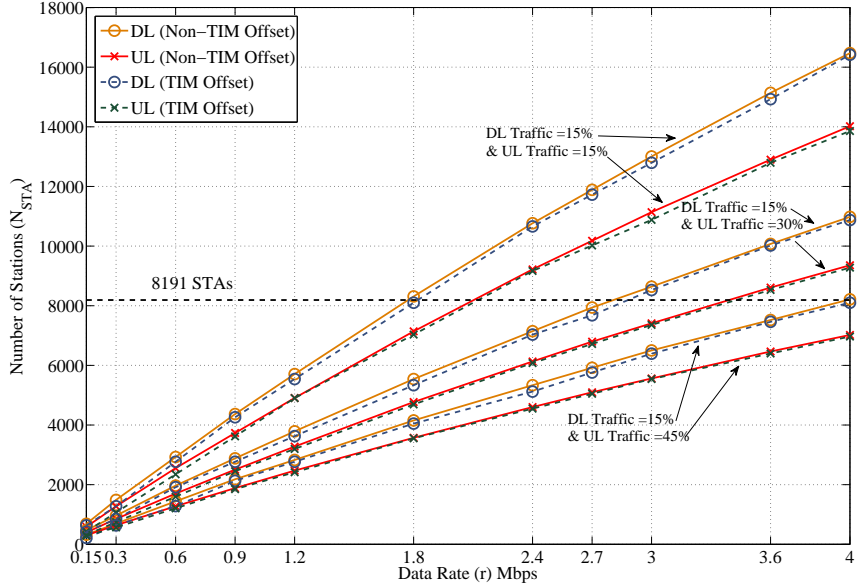}
\caption{Maximum number of supported STAs by an IEEE 802.11ah WLAN in function of different PHY Layer data rates, traffic patterns and signalling modes.}
\label{fig:theoretical}
\end{figure}
\end{center}

Figure \ref{fig:theoretical} shows the maximum number of supported STAs in an IEEE 802.11ah WLAN depending on the data rate. Results reflect some cases in which the maximum number of STAs supported by the network surpasses the 8191 STAs. Two traffic patterns (DL=15\% UL=15\% and DL=15\% UL=30\%) are able to support the 8191 stations. While the first one achieves this goal with r=2.4 Mbps, the second one does it with r=3.6 Mbps. Traffic pattern with DL=15\% and UL=45\% is only capable of supporting 6967 stations with r=4 Mbps. Besides, the differences between the Non-TIM and TIM Offset signalling modes are also shown. Results reflect a higher value in the maximum number of supported STAs for the Non-TIM Offset mode in all scenarios considered.

\vspace{-1cm}
\subsection{Packet Delivery Ratio and Network Efficiency}

In this section we evaluate the PDR and the Network Efficiency $(\eta_{\psi})$ in a specific scenario where the number of STAs is large. These two figures of merit are calculated as:

$$
\text{PDR}_{\psi} = \left(\frac{\text{Packets Delivered}_{\psi}}{\text{Packets Generated}_{\psi}}\right)
$$

$$
\eta_{\psi} = \left(\frac{\text{Packets Delivered}_{\psi}}{\text{Channel Capacity}_{\psi}}\right)
$$

with Channel Capacity$_{\psi}$, i.e. the maximum number of DL/UL data packets that the network could be able to deliver in a simulation, and is computed as: $$\text{Channel Capacity}_{\psi}=N_{\psi} \cdot N_{\DTIM}$$

We assume a data rate of $r=1.8$ Mbps, and set the number of STAs to $7140$ for Scenario A, $4770$ for Scenario B and $3571$ for Scenario C. It is worth noting that these values are the maximum number of STAs supported when the Non-TIM Offset signalling mode is used. 

\begin{figure}[t!]
\centering
\includegraphics[scale=0.35]{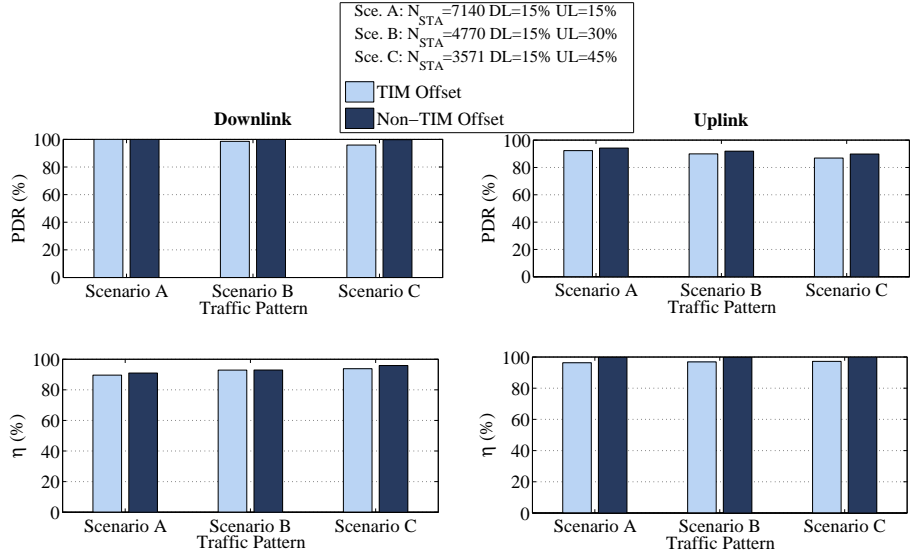}
\caption{Packet Delivery Ratio and Network Efficiency versus Different Transmission Rates and for TIM and Non-TIM Offset Schemes}
\label{fig:traffic}
\end{figure}

The results shown in Figure \ref{fig:traffic} reflect a high value of PDR for both modes: more than 90\% in any studied case and always better when using the Non-TIM Offset. Nevertheless, the difference in terms of PDR, has a minimum value of 1.9\% and a maximum value of 3\%. Therefore, these results reflect the efficient operation of an IEEE 802.11ah WLANs when the number of STAs is very high. As one can observe, the results of the Network Efficiency achieved are always above 89\%.

\subsection{Energy Consumption}

\begin{figure}[h!]
\centering
\includegraphics[scale=0.38]{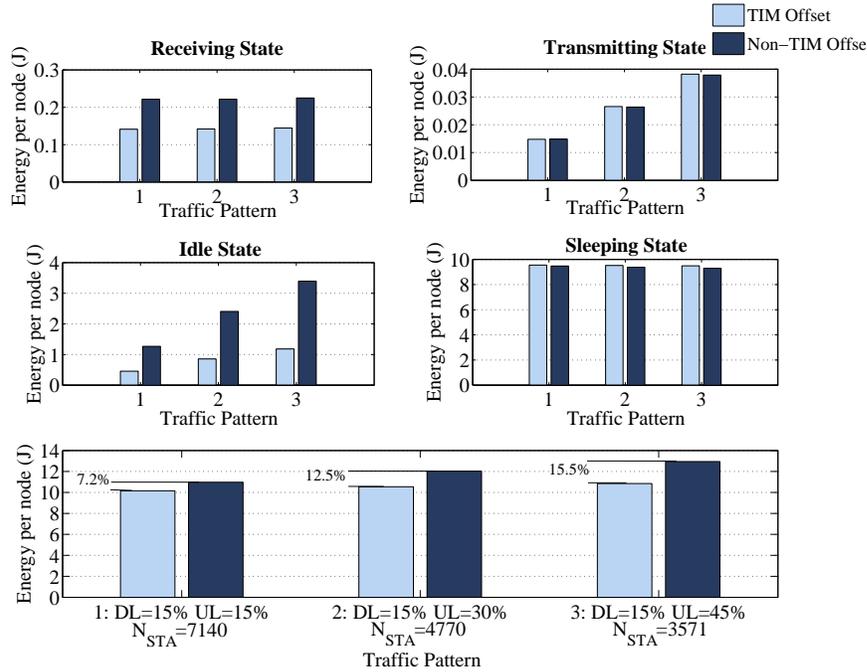}
\caption{Energy Consumed per Node ( \small{\textit{Tx mode = 1400 mW; Rx mode = 900 mW; Id mode = 700 mW; Sl = 60 mW \cite{orinoco}}})}
\label{fig:energy}
\end{figure}

Finally, the last figure of merit evaluated is the energy consumed by the STAs. The same scenarios and number of STAs as in previous case are considered. 

The results are shown in Figure \ref{fig:energy} and reflect one of the major issues that IEEE 802.11ah amendment aims for: the reduction of the energy consumption through the use of low power mechanisms, as described in Section \ref{Sec:MAC}. The energy consumed is divided in four states:
\begin{itemize}
\item \textbf{Receiving}: STAs in PSM which have not entered into a Long Sleeping Period must listen to all the DTIM beacons. If a STA is signalled in a DTIM beacon with downlink data or it has data to transmit, it will also listen to its corresponding TIM beacon. A STA receiving a data packet, a CTS or an ACK is also in receiving state. Overhearing of packets addressed to other STAs is also affecting the time a STA is in receiving mode.
	\item \textbf{Idle}: It is referred to Backoff periods and interframe spaces such as SIFS and DIFS.
	\item \textbf{Transmitting}: When STAs have to transmit certain frames both in the downlink (PS-Poll, ACK) and the uplink (RTS, DATA) communications.
	\item \textbf{Sleeping}: When STAs switch off their radio module. 
\end{itemize}

Although in both signalling modes, Non-TIM and TIM Offset, nodes remain the majority of the time in sleeping state, the major differences of consumption are reflected in the receiving and in the idle state. The Non-TIM Offset mechanism consumes more energy in these two states. 

As presented in Section \ref{Sec:MAC}, the beacons used in Non-TIM mechanism are larger due to the increase of signalling data. Hence, STAs receive longer beacons, which results in that they have to remain more time in the receiving state. On the other hand, the TIM Offset mode divides each DL$\setminus$UL time period in different page segments. This time division allows to distribute the STAs among more segments, compared to the Non-TIM signalling mode. Hence, STAs, that are only allowed to transmit inside their assigned segment, remain asleep during the rest of them. For that reason, their consumption in the idle state is reduced.

These two facts are reflected in the total energy consumed per STA shown in Figure \ref{fig:energy}. The TIM Offset mode achieves lower values of total energy consumed than the Non-TIM Offset mode. For the first traffic pattern simulated the reduction corresponds to 7.2\%. For the second pattern, the gain achieved is 12.5\%, while for the third one, the energy consumed is reduced in 15.5\%.


\section{Conclusions} \label{Sec:Conclusions}

In this paper, a theoretical model to compute the maximum number of STAs supported in an IEEE 802.11ah WLAN is presented.  Besides, the comparison between the Non-TIM and TIM Offset signalling modes for the new IEEE 802.11ah amendment has been evaluated. In the different scenarios evaluated, simulation results have shown that these mechanisms achieve a good PDR for both Downlink and Uplink traffic when the number of STAs is large. 

From the PDR values obtained in the simulations, the theoretical model can be considered as a valid upper bound for that network parameter. It is also demonstrated the better behaviour of Non-TIM Offset signalling mode in terms of the maximum number of STAs supported. 

Besides, our simulations show that, in the considered scenarios, STAs remain in the sleeping mode more than $98$\% of the time. As a consequence, the energy consumed by STAs will be very low, what confirms the suitability of the presented protocol for battery-powered sensor and actuator networks. In detail, the TIM Offset mode shows a lower energy consumption compared to the Non-TIM one. However, the PDR achieves lower values when using the TIM Offset mode, with a maximum difference of 3\%. 

Some areas for future work have been detected. For instance, the study of the effects related to the presence of hidden terminals, non-TIM and Unscheduled STAs or traffic differentiation mechanisms, in addition to the existence of network association/disassociation and long sleeping mechanisms. 


\section*{Acknowledgements}

This work has been partially supported by the Spanish Government under projects TEC2012-32354 and IPT-2012-1028-120000 and by the Catalan Government (SGR2009\#00617).

%
\bibliographystyle{IEEEtran}
\bibliography{Bib}

\end{document}